\documentstyle[12pt]{article}
\newcommand{\definition}{\stackrel{\rm d}{\equiv}}
\begin{document}
\begin{titlepage}
\begin{flushright}
March 1995
\vspace{3cm}\\
\end{flushright}
\begin{center}
{\bf\Huge Higher Derivatives and Canonical Formalism}
\vspace{4cm}\\
{\Large Shinji HAMAMOTO}
\vspace{1cm}\\
{\it Department of Physics, Toyama University \\
Toyama 930, JAPAN}
\vspace{3cm}\\
\end{center}
\begin{abstract}
A canonical formalism for higher-derivative theories is presented
on the basis of Dirac's method for constrained systems.
It is shown that this formalism shares a path integral expression
with Ostrogradski's canonical formalism.
\end{abstract}
\end{titlepage}

\section{Introduction}

Higher-derivative theories appear in various scenes of physics:${}^{1),2)}$
higher derivative terms occur as quantum corrections;
nonlocal theories, e.g. string theories, are essentially higher-derivative the
ories;
Einstein gravity supplemented by curvature-squared terms has attracted attenti
on because of its renormalizability.${}^{3)}$

A canonical formalism for higher-derivative theories was developed by Ostrogra
dski about one and a half centuries ago.${}^{4)}$
Though being self-consistent, his formulation looks different from the convent
ional canonical formalism.

The purpose of the present paper is to interpret the Ostrogradski's formulatio
n in the framework of the ordinary constrained canonical formalism.
We show that the path integral expression for the ordinary constrained canonic
al formalism is the same as that for the Ostrogrdski formalism.

In \S 2 we review the Ostrogradski's formulation.
In \S 3 we give another canonical formulation for higher-derivative theories b
ased on the usual method of Dirac for constrained systems.${}^{5)}$
In \S 4 it is shown that the two formulations give the same path integral form
ulae.
Section 5 gives summary.

\section{Ostrogradski's canonical formalism}

We conside a generic Lagrangian which contains up to $N$th derivative of coord
inate $q(t)$:
\begin{equation}
L = L \left( q^{(0)}, q^{(1)}, \cdots , q^{(N)} \right) ,
\end{equation}
where
\begin{equation}
q^{(i)} \definition \frac{{\rm d}^{i}}{{\rm d}t^{i}}q
\makebox[1cm]{} ( i = 0, 1, \cdots , N ) .
\end{equation}
The Euler-Lagrange equation is
\begin{equation}
\frac{\delta S}{\delta q} \definition
\sum_{i=0}^{N} \left( - \frac{{\rm d}}{{\rm d}t} \right) ^{i}
\frac{\partial L}{\partial q^{(i)}} = 0 .
\end{equation}

The canonical formalism of Ostrogradski is the following.${}^{4)}$
For $i = 0, \cdots , N-1$ we regard $q^{(i)}$ as independent coordinates $q^{i
}$ :
\begin{eqnarray}
q^{(i)}
& \longrightarrow &
q^{i} \makebox[1cm]{} ( i = 0, 1, \cdots , N-1 ) , \\
L \left( q^{(0)}, q^{(1)}, \cdots , q^{(N-1)}, q^{(N)} \right)
& \longrightarrow &
L \left( q^{0}, q^{1}, \cdots , q^{N-1}, \dot{q}^{N-1} \right) .
\end{eqnarray}
The momentum $p_{N-1}$ conjugate to $q^{N-1}$ is defined as usual by
\begin{equation}
p_{N-1} \definition \frac{\partial L}{\partial \dot{q}^{N-1}}
\left( q^{0}, q^{1}, \cdots , q^{N-1}, \dot{q}^{N-1} \right) .
\end{equation}
Here and hereafter we assume that the Lagrangian is nondegenerate.
This means that the relation (6) can be inverted to give $\dot{q}^{N-1}$ as a
function of $q^{i} \; ( i = 0, \cdots , N-1 )$ and $p_{N-1}$ :
\begin{equation}
\dot{q}^{N-1} = \dot{q}^{N-1}
                \left( q^{0}, q^{1}, \cdots , q^{N-1}, p_{N-1} \right) .
\end{equation}
The Hamiltonian is {\it defined} by
\begin{eqnarray}
H_{{\rm O}} & =           &
H_{{\rm O}} \left( q^{0}, q^{1}, \cdots , q^{N-1}\, ; \,
             p_{0}, p_{1}, \cdots, p_{N-1} \right) \nonumber \\
      & \definition &
\sum_{i=0}^{N-2} p_{i}q^{i+1} + p_{N-1}\dot{q}^{N-1}
- L \left( q^{0}, q^{1}, \cdots , q^{N-1}, \dot{q}^{N-1} \right) .
\end{eqnarray}
This definition is different from the usual Legendre transformation.
Note that for $i = 0, \cdots , N-2$, the momenta $p_{i}$ are {\it not} defined
 through relations like (6), but introduced just as independent canonical vari
ables.
The canonical equations of motion are
\begin{eqnarray}
\dot{q}^{i} & = & \frac{\partial H_{{\rm O}}}{\partial p_{i}} , \\
\dot{p}_{i} & = & \mbox{} \! - \frac{\partial H_{{\rm O}}}{\partial q^{i}} .
\end{eqnarray}
Eq.(9) with $i = 0, \cdots , N-2$ gives
\begin{equation}
\dot{q}^{i} = q^{i+1} \makebox[1cm]{} ( i = 0, 1, \cdots , N-2 ).
\end{equation}
Under the assumption of nondegeneracy, Eq.(9) with $i = N-1$ reproduces the de
finition (6).
Eq.(10) gives the following equations:
\begin{equation}
\left\{
\begin{array}{rcl}
\dot{p}_{0} & = & \displaystyle
                  \frac{\partial L}{\partial q^{0}} , \vspace{2mm}\\
\dot{p}_{i} & = & \displaystyle
                  \mbox{} \! - p_{i-1} + \frac{\partial L}{\partial q^{i}}
                  \makebox[1cm]{} ( i = 1, \cdots , N-1 ) .
\end{array}
\right.
\end{equation}
{}From Eqs.(12), (11) and (6) we regain the Euler-Lagrange equation (3).

\section{Constrained canonical formalism}

It has been seen that the Ostrogradski formalism gives special treatment to 
the highest derivative $q^{(N)}$.
Is it possible to treat the highest derivative and the lower derivatives on an
 equal footing?
This is the subject of the present section.

To treat all the derivatives equally, we introduce Lagrange multipliers 
$\lambda _{i}\; ( i = 0, \cdots , N-1 )$ and start from the folllowing 
equivalent Lagrangian:
\begin{equation}
L^{*} \definition
L \left( q^{0}, q^{1}, \cdots , q^{N} \right)
+ \sum_{i=0}^{N-1} \lambda_{i} \left( \dot{q}^{i} - q^{i+1} \right) .
\end{equation}
The Euler-Lagrange equations
\begin{equation}
\left\{
\begin{array}{rcl}
\displaystyle \frac{\delta L^{*}}{\delta q^{i}}       & = &
0 \makebox[1cm]{} ( i = 0, \cdots , N ) , \vspace{2mm}\\
\displaystyle \frac{\delta L^{*}}{\delta \lambda_{i}} & = &
0 \makebox[1cm]{} ( i = 0, \cdots , N-1 )
\end{array}
\right.
\end{equation}
give
\begin{equation}
\left\{
\begin{array}{l}
\displaystyle \sum_{i=0}^{N} \left( - \frac{{\rm d}}{{\rm d}t} \right) ^{i}
\frac{\partial L}{\partial q^{i}} = 0 , \vspace{2mm}\\
\displaystyle \dot{q}^{i} = q^{i+1} \makebox[1cm]{} ( i = 0, \cdots , N-1 ) ,
\end{array}
\right.
\end{equation}
which are equivalent to Eq.(3).

Since the Lagrangian $L^{*}$ describes a constrained system, we follow a way o
f Dirac.${}^{5)}$
The conjugate momenta defined by
\begin{equation}
\left\{
\begin{array}{rcl}
p_{i}    & \definition &
\displaystyle \frac{\partial L^{*}}{\partial \dot{q}^{i}}
\makebox[1cm]{} ( i = 0, \cdots , N ) , \vspace{2mm}\\
\pi ^{i} & \definition &
\displaystyle \frac{\partial L^{*}}{\partial \dot{\lambda}_{i}}
\makebox[1cm]{} ( i = 0, \cdots , N-1 )
\end{array}
\right.
\end{equation}
provide the following primary constraints:
\begin{eqnarray}
\phi _{i} & \definition & p_{i} - \lambda _{i} \: \approx \: 0
\makebox[1cm]{} ( i = 0, \cdots , N-1 ) , \\
\phi _{N} & \definition & p_{N} \makebox[1cm]{} \!\!\! \approx \: 0 , \\
\psi ^{i} & \definition & \pi ^{i} \makebox[1cm]{} \! \approx \: 0
\makebox[1cm]{} ( i = 0, \cdots , N-1 ) .
\end{eqnarray}
The consistency of these constraints under their time developments requires a
secondary constraint
\begin{equation}
\psi ^{N} \definition \frac{\partial L}{\partial q^{N}} - \lambda _{N-1}
\approx 0 .
\end{equation}
When the system is nondegenerate, all these constraints
\begin{equation}
\Phi _{\alpha} \definition
\left( \phi _{0}, \cdots , \phi _{N} \, ;
\, \psi ^{0}, \cdots , \psi ^{N} \right)
\end{equation}
form a set of second-class constraints:
the determinant of the Poisson brackets between these constraints
\begin{equation}
\det \left( \left[ \Phi _{\alpha}, \Phi _{\beta} \right] _{{\rm P}} \right)
= \left( \frac{\partial ^{2}L}{\partial q^{N2}} \right) ^{2}
\end{equation}
is not zero when
\begin{equation}
\frac{\partial ^{2}L}{\partial q^{N2}} \neq 0 .
\end{equation}
The Dirac brackets between the canonical variables $\left( q^{i}, \lambda _{i}
 \, ; \, p_{i}, \pi ^{i} \right)$ are calculated to be
\begin{equation}
\left\{
\begin{array}{rllll}
\left[ q^{i}, p_{i} \right] _{{\rm D}} & = &
\left[ q^{i}, \lambda _{i} \right] _{{\rm D}} & = & 1 , \vspace{2mm}\\
\left[ q^{N}, p_{i} \right] _{{\rm D}} & = &
\left[ q^{N}, \lambda _{i} \right] _{{\rm D}} & = &
\displaystyle - \left( \frac{\partial ^{2}L}{\partial q^{N2}} \right) ^{-1}
\frac{\partial ^{2}L}{\partial q^{i} \partial q^{N}} , \vspace{2mm}\\
\left[ q^{N-1}, q^{N} \right] _{{\rm D}} & = &
\displaystyle \left( \frac{\partial ^{2}L}{\partial q^{N2}} \right) ^{-1} ,
& & \vspace{2mm}\\
{\rm The}\ {\rm others} & = & 0 ,& &
\end{array}
\right.
\end{equation}
\[
( i = 0, \cdots , N-1 ) .
\]
The Hamiltonian is given by
\begin{equation}
H_{{\rm D}} = - L \left( q^{0}, \cdots , q^{N} \right) +
\sum_{i=0}^{N-1} \lambda _{i}q^{i+1} .
\end{equation}
This is obtained by performing the ordinary Legendre transformation on the Lag
rangian $L^{*}$ of Eq.(13) and by regarding all the constraints $\Phi _{\alpha
} \approx 0$ as strong equalities $\Phi _{\alpha} = 0$.
Eqs.(25) and (24) allow us to obtain the canonical equations of motion.
The independent equations are
\begin{equation}
\left\{
\begin{array}{llll}
\dot{q}^{i} & = & q^{i+1} & ( i = 0, \cdots , N-1 ) , \vspace{2mm}\\
\dot{\lambda}_{0} & = &
\displaystyle \frac{\partial L}{\partial q^{0}} , & \vspace{2mm}\\
\dot{\lambda}_{i} & = &
\displaystyle \frac{\partial L}{\partial q^{i}} - \lambda _{i-1} &
( i = 1, \cdots , N-1 ) , \vspace{2mm}\\
0 & = & \displaystyle \frac{\partial L}{\partial q^{N}} - \lambda _{N-1} ,
&
\end{array}
\right.
\end{equation}
which are seen to be equivalent to the Euler-Lagrange equations (15).

\section{Path integrals}

We have presented two canonical formalisms for higher-derivative theories, the
 Ostrogradski's one and the constrained one.
It has been seen that though looking different from each other, they give an e
quivalent set of equations of motion.
In this section we show that the two formalisms are completely equivalent to e
ach other by comparing path integral expressions for them.

The Ostrogradski formalism of \S 2 gives the following path integral expressio
n:
\begin{equation}
Z_{{\rm O}} = \int \prod_{i=0}^{N-1}\left( {\cal D}q^{i}{\cal D}p_{i} \right)
\exp \left\{ i\int {\rm d}t \left[
\sum_{i=0}^{N-1} p_{i}\dot{q}^{i} -
H_{{\rm O}} \left( q^{0}, \cdots , q^{N-1}\, ; \, p_{0}, \cdots , p_{N-1} \right)
\right] \right\} ,
\end{equation}
where the Hamiltonian $H_{\rm O}$ is given by Eq.(8).
Integrations with respect to $p_{i}\; ( i = 0, \cdots , N-2 )$ offer a factor
$\prod _{i=0}^{N-2} \delta \left( \dot{q}^{i} - q^{i+1} \right)$.
We can further integrate with respect to $q^{i} \; \left( i = 1, \cdots , N-1
\right)$, obtaining
\begin{equation}
Z_{{\rm O}} = \int {\cal D}q{\cal D}p_{N-1}
\exp \left\{ i\int {\rm d}t \left[
p_{N-1}q^{(N)} -
\hat{H} \left( q, \dot{q}, \cdots , q^{(N-1)}, p_{N-1} \right)
\right] \right\} ,
\end{equation}
where
\begin{equation}
\hat{H} \definition p_{N-1}\dot{q}^{N-1}
- L \left( q, \dot{q}, \cdots , q^{(N-1)}, \dot{q}^{N-1} \right) .
\end{equation}
In Eq.(29), $\dot{q}^{N-1}$ is a function of $q^{(i)} \; \left( i = 0, \cdots
, N-1 \right)$ and $p_{N-1}$, which is obtained by replacing $q^{i}$ with $q^{
(i)}$ in Eq.(7).
Path integral expression for the case of the constrained canonical formalism o
f \S 3 is
\begin{eqnarray}
Z_{{\rm D}} & = &
\int \prod_{i=0}^{N} \left( {\cal D}q^{i}{\cal D}p_{i} \right)
     \prod_{i=0}^{N-1} \left( {\cal D}\lambda _{i}{\cal D}\pi ^{i} \right)
     \prod_{i=0}^{N} \left( \delta (\phi _{i})\delta (\psi ^{i}) \right)
     \left| \frac{\partial ^{2}L}{\partial q^{N2}} \right| \nonumber \\
& & \makebox[5mm]{} \times \exp \left\{
i\int {\rm d}t \left[ \sum_{i=0}^{N} p_{i}\dot{q}^{i} +
\sum_{i=0}^{N-1} \pi {i}\dot{\lambda}_{i} -
H_{\rm D} \left( q^{0}, \cdots , q^{N} \, ; \, \lambda _{0}, \cdots , \lambda
_{N-1}
\right) \right] \right\} , \nonumber \\
& &
\end{eqnarray}
where the constraints $\left( \phi _{i}, \psi ^{i} \right) \; ( i = 0, \cdots
, N )$ are given by Eqs.(17) -- (20), and the Hamiltonian $H_{\rm D}$ by Eq.(2
5).
Integrations with respect to $p_{N}, \; \pi ^{i} \; ( i = 0, \cdots , N-2 ), \
; \lambda _{i} \; ( i = 0, \cdots , N-2 ), \; p_{i} \; ( i = 0, \cdots , N-2 )
$ and $q^{i} \; ( i = 1, \cdots , N )$ give
\begin{eqnarray}
Z_{{\rm D}} & = &
\int \prod_{i=0}^{N}{\cal D}q^{i}\prod_{i=0}^{N-1}{\cal D}p_{i}
\delta \left( \frac{\partial L}{\partial q^{N}} - p_{N-1} \right)
\left| \frac{\partial ^{2}L}{\partial q^{N2}} \right| \nonumber \\
& & \makebox[2cm]{} \times \exp \left\{
i\int {\rm d}t \left[
\sum_{i=0}^{N-1} p_{i}\left( \dot{q}^{i} - q^{i+1} \right) +
L \left( q^{0}, \cdots , q^{N} \right) \right] \right\} \nonumber \\
& = &
\int \prod_{i=0}^{N}{\cal D}q^{i}{\cal D}p_{N-1}
\delta \left( \frac{\partial L}{\partial q^{N}} - p_{N-1} \right)
\left| \frac{\partial ^{2}L}{\partial q^{N2}} \right|
\prod_{i=0}^{N-2}\delta \left( \dot{q}^{i} - q^{i+1} \right) \nonumber \\
& & \makebox[2cm]{} \times \exp \left\{
i\int {\rm d}t \left[
p_{N-1}\left( \dot{q}^{N-1} - q^{N} \right) +
L \left( q^{0}, \cdots , q^{N} \right) \right] \right\} \nonumber \\
& = &
\int {\cal D}q{\cal D}p_{N-1}
\exp \left\{ i\int {\rm d}t \left[
p_{N-1}\left( q^{(N)} - q^{N} \right) +
L \left( q, \dot{q}, \cdots , q^{(N-1)}, q^{N} \right) \right] \right\} .
\nonumber \\
& &
\end{eqnarray}
In the last line of Eq.(31), $q^{N}$ is a function of $q^{(i)} \; ( i = 0, \cdots , N-1 )$ and $p_{N-1}$ defined by inverting the relation
\begin{equation}
p_{N-1} = \frac{\partial L}{\partial q^{N}}
\left( q, \dot{q}, \cdots , q^{(N-1)}, q^{N} \right) .
\end{equation}
That means $q^{N}$ is nothing but $\dot{q}^{N-1}$ in Eq.(29).
Putting $q^{N} = \dot{q}^{N-1}$ in Eq.(31) shows that the path integral $Z_{{\
rm D}}$ is the same as $Z_{{\rm O}}$ given by Eq.(28)
\begin{equation}
Z_{\rm D} \equiv Z_{\rm O} .
\end{equation}

\section{Summary}

We have presented a canonical formalism for higher-derivative theories based o
n the usual method of Dirac for constrained systems.
It has been shown that this formalism shares a path integral expression with t
he Ostrogradski's one.
We thus have laid the foundation of the ordinary canonical formalism for the O
strogradski's formulation.

\section*{Acknowledgments}

The author would like to thank Minoru Hirayama, Shinobu Hosono and Hitoshi Yam
akoshi for discussions.


\begin{thebibliography}{1}
\bibitem{} D. A. Eliezer and R. P. Woodard, Nucl. Phys. B325 (1989) 389.
\bibitem{} J. Z. Simon, Phys. Rev. D41 (1990) 3720.
\bibitem{} K. S. Stelle, Phys. Rev. D16 (1977) 953.
\bibitem{} M. Ostrogradski, Mem. Ac. St. Petersbourg VI4 (1850) 385.
\bibitem{} P. A. M. Dirac, {\it Lectures on Quantum Mechanics} (Yeshiva Univer
sity Press, New York, 1964).
\end{thebibliography}
\end{document}